\documentclass[a4paper,12pt]{article}
\usepackage{amsfonts}
\usepackage[usenames]{color}
\usepackage{amsmath}
\usepackage{amssymb}

\newcommand{\be}{\begin{equation}}
\newcommand{\ee}{\end{equation}}
\newcommand{\ba}{\begin{eqnarray}}
\newcommand{\ea}{\end{eqnarray}}
\newcommand{\pd}{\partial}

\title{\bf \Large Strings in extremal BTZ black holes}
\author{Jamie Parsons\footnote{J.D.Parsons@durham.ac.uk} {} and Simon
  F. Ross\footnote{S.F.Ross@durham.ac.uk} \\ \\ \small \sl Centre for
  Particle Theory \& Department of 
Mathematical Sciences,
\\[-1.5mm]
\small \sl Durham Univerity, South Road, Durham DH1 3LE, United Kingdom. \\
}

\begin{document}

\begin{titlepage}
\maketitle
\begin{picture}(0,0)(0,0)
\put(350, 303){DCPT-09/05} 
\end{picture}
\vspace{-36pt}

\begin{abstract}
  We study the spectrum of the worldsheet theory of the bosonic closed
  string in the massless and extremal rotating BTZ black holes. We use
  a hyperbolic Wakimoto representation of the $SL(2,\mathbb{R})$
  currents to construct vertex operators for the string modes on these
  backgrounds. We argue that there are tachyons in the twisted sector,
  but these are not localised near the horizon. We study the relation
  to the null orbifold in the limit of vanishing cosmological
  constant. We also discuss the problem of extending this analysis to
  the supersymmetric case.
\end{abstract}
\thispagestyle{empty}
\setcounter{page}{0}
\end{titlepage}

\section{Introduction}
\label{intro}

The Banados-Teitelboim-Zanelli (BTZ) black
hole~\cite{Banados:1992wn,Banados:1992gq} is a very useful laboratory
for exploring aspects of black holes and geometry in a simplified
setting, as the geometry is simply an orbifold of AdS$_3$. In
particular, if we consider an AdS$_3 \times S^3$ solution in string
theory supported by NS-NS fluxes, the spectrum of perturbative string
theory on the BTZ black hole can easily be determined using standard
orbifold techniques. This problem has been extensively investigated
for non-extremal BTZ black holes, which are orbifolds of AdS$_3$ along
a hyperbolic
generator~\cite{Natsuume:1996ij,Satoh:1997xe,Hemming:2001we,Martinec:2002xq,Hemming:2002kd,Rangamani:2007fz}. The
twisted sector states for the orbifold are obtained using a
parafermionic representation of the current algebra and a twist
operator construction which are based on earlier work on the long
string sectors of global AdS$_3$~\cite{Argurio:2000tb} and closely
related to the work on the elliptic orbifold
in~\cite{Martinec:2001cf}. The extremal black holes, which are
orbifolds along a parabolic generator, are less studied (although the
winding sectors were considered in~\cite{Troost:2002wk}, where the
relevance of the hyperbolic Wakimoto representation of the currents
used in~\cite{Satoh:1997xe} was recognized). Since the extremal cases
are a different class of orbifolds from the non-extremal cases, they
will require separate consideration. From the worldsheet point of
view, the parafermionic representation used in the non-extremal case
are no longer appropriate, and we need to find a new representation of
the vertex operators which diagonalises the angular momentum in the
extremal BTZ black hole.

This paper addresses this problem, finding an explicit set of vertex
operators for the untwisted and twisted sectors for the bosonic string
on the zero-mass and extremal rotating BTZ black holes. We consider an
AdS$_3 \times S^3$ geometry supported by NS-NS flux, corresponding to
an F1-NS5 system compactified on a Ricci-flat internal manifold. The
world-sheet theory is a CFT with an $\widehat{SL(2,\mathbb{R})}_k
\times \widehat{SU(2)}_k$ current algebra, with the level $k$ being
set by the NS-NS flux. We will discuss the bosonic string in detail;
the problem of extending our analysis to the superstring will be
discussed at the end. We want to work in a parabolic basis for
$SL(2,\mathbb{R})$, which diagonalises the combination of generators
corresponding to the momentum along the compact circle in zero-mass
BTZ.  Here we show that the hyperbolic Wakimoto representation
introduced in~\cite{Satoh:1997xe}, provides an appropriate
representation of the $\widehat{SL(2,\mathbb{R})}_k$ current
algebra. This representation has the advantage that the expression for
the vertex operators is more explicit than in the parafermionic
representation used in the non-extremal case.

We apply this calculation of the spectrum to study the tachyons in
this background. Tachyons in the non-extremal non-rotating BTZ black
hole were studied in~\cite{Rangamani:2007fz}, as an explicit example
of the kind of quasi-localised closed string tachyons discussed in
\cite{Adams:2005rb}.\footnote{As BTZ arises as the near-horizon limit
  of a charged black string, it is directly related to the examples
  discussed in \cite{Horowitz:2005vp, Ross:2005ms}. Other examples
  with a quasi-localised tachyon include
  \cite{McGreevy:2005ci,Horowitz:2006mr}.} The idea is that if we
consider string theory compactified on a circle, when the size of the
circle is smaller than the string length $\ell_s$, there are tachyonic
winding modes. If the size of this circle varies over some base space,
one heuristically expects a tachyon which is confined to the region
where the size of the circle $\le \ell_s$. It was found
in~\cite{Rangamani:2007fz} that there is a tachyon in the twisted
sector NS-NS ground state if the size of the circle at the black hole
horizon is smaller than the string scale, $\sqrt{k} r_+ \leq
\ell_s$. However, this tachyon is found not to be localised in the
near-horizon region, due to the coupling to the NS-NS field. As the
zero-mass BTZ black hole is the limit as $r_+ \to 0$, we would expect
that in this case, the NS-NS ground state in twisted sectors should
always be tachyonic, and using the explicit representation of the
spectrum we obtain, that is indeed what we find. We also extend the
analysis to the rotating case, showing that tachyons arise in the
twisted sectors if $\sqrt{k} r_+ \leq 2 \ell_s$, as in the previous
discussion of the non-rotating non-extremal case. The study of
tachyons in the $M=0$ BTZ black hole has a couple of advantages over
the previous non-extremal case. Firstly, the expressions for the
vertex operators in the Wakimoto representation are more explicit than
the parafermionic representation used in the previous case. Secondly,
the geometry has a causal Killing vector everywhere, so issues of
tachyon condensation could be addressed in the $M=0$ BTZ black hole
without having to deal with the complications of studying the
behaviour on a time-dependent background geometry in the region behind
the horizon. We will not address the question of the condensation of
the tachyon, which remains a challenging direction for future work.

In section 2, we review aspects of the $M=0$ BTZ black hole. In
section 3, we introduce the hyperbolic Wakimoto representation of the
current algebra, and use it to construct vertex operators for the
untwisted sector states. We then introduce a twist operator enforcing
the orbifold condition, and use it to obtain the twisted sector vertex
operators. We discuss the condition for a tachyon to exist in the
spectrum, and argue that the NS-NS ground states in the twisted sector
are tachyonic, as expected. In section 4, we discuss the flat space
limit, taking $k \to \infty$ while focusing on the neighbourhood of
the singularity. In this limit, the zero mass BTZ black hole reduces
to the null orbifold of flat space~\cite{Steif:1995zm,Simon:2002ma}.

In section 5, we extend the analysis to the extremal rotating BTZ black
hole. We show that the relevant orbifold action is chiral, with the
action on left-movers the same as for the zero-mass black hole while
the action on right-movers is the same as for the non-zero mass black
hole. We can thus construct appropriate vertex operators by combining
the previous results for these two cases. We show that the resulting
set of vertex operators for twisted sectors is mutually local, and
argue that a tachyon appears when $\sqrt{k} r_+ \leq 2 \ell_s$, as
expected. 

In section 6, we discuss the extension of our results to the
superstring. The main open problem is to find a representation of the
spin fields which diagonalises the action of the spacetime angular
momentum. Without such a representation, we cannot explicitly
construct vertex operators corresponding to the modes which survive
the orbifold projection in the NS-R and R-R sectors. The final section
summarises the paper and considers possible routes for further
investigation.

\section{AdS$_3$ worldsheet theory}

Bosonic string theory on $AdS_3$ is described by the
$SL(2,\mathbb{R})$ WZW model with action
\be
S_{WZW} = \frac{k}{8\pi\alpha'}\int d^2\sigma Tr(g^{-1}\pd_a g g^{-1} \pd^a g) +\frac{ik}{12\pi}\int Tr(g^{-1}dg\wedge g^{-1}dg\wedge g^{-1}dg),
\ee
where $g$ is the $SL(2,\mathbb{R})$ group element, $k$ is the level,
and the second integral runs over a three-dimensional manifold whose
boundary is the two-dimensional worldsheet (see
e.g.~\cite{Maldacena:2000hw} for a nice discussion of strings on
AdS$_3$). Henceforth we will set $\alpha'=1$, so we work in units of
the string length. The AdS length scale is then $\ell = \sqrt{k}$. The
WZW model is invariant under the action
\be
g(z,\bar{z}) \to \omega(z)g(z,\bar{z})\bar{\omega}(\bar{z})^{-1},
\ee
which leads to a set of conserved world-sheet currents,
\be \label{J} \mathcal{J}(z) = \mathcal{J}_{a}\tau^{a} =
-\frac{k}{2}\partial g g^{-1}, \qquad \bar{\mathcal{J}}(\bar{z}) =
\bar{\mathcal{J}}_{a}\tau^{a} = -\frac{k}{2}g^{-1}\bar{\partial}g,
\ee
where the generators of $SL(2,\mathbb{R})$ are given by
\begin{eqnarray} \label{slgens}
 \tau^{1} &=& \frac{i}{2} \sigma^{3} = \frac{1}{2}\left( \begin{array}{cc}
                            i & 0 \\
                            0 & -i
                           \end{array} \right), \\
\tau^{2} &=& \frac{i}{2} \sigma^{1} = \frac{1}{2}\left( \begin{array}{cc}
                            0 & i \\
                            i & 0
                           \end{array} \right), \nonumber \\
\tau^{3} &=& \frac{1}{2} \sigma^{2} = \frac{1}{2}\left( \begin{array}{cc}
                            0 & -i \\
                            i & 0
                           \end{array} \right). \nonumber
\end{eqnarray}
These satisfy the algebra
\be \label{commreln}
[\tau^a,\tau^b] = i\epsilon^{ab}_{\ \ c}\tau^c,
\ee
where $\epsilon^{123} = 1$, and the index is lowered with the metric
$\eta_{ab} = {\mbox diag}(1,1,-1)$.  It will later be convenient to
introduce the combinations $\tau^\pm = \tau^3 \pm \tau^2$. Note that
this choice of notation doesn't correspond to the usual conventions
for raising and lowering operators: we have the commutation relations
$[\tau^1, \tau^+] = -i \tau^+$, $[\tau^1, \tau^-] = i \tau^-$,
$[\tau^+, \tau^-] = 2i \tau^1$.

Using the Ward identity
\be \label{Ward} 
i\delta A (w, \bar{w}) = \oint_{w}\frac{dz}{2\pi i}\epsilon_{a}\mathcal{J}^{a}A (w, \bar{w}) + \oint_{w}\frac{d\bar{z}}{2\pi i}\bar{\epsilon}_{a}\bar{\mathcal{J}}^{a}A (w, \bar{w})
\ee 
we can find the OPEs for the currents. They are
\be
\mathcal{J}^{b}\mathcal{J}^{c}=\frac{i\epsilon^{bc}_{\ \ a}\mathcal{J}^{a}}{z-w}
+ \frac{\frac{k}{2}\eta^{bc}}{(z-w)^{2}}, \qquad
\bar{\mathcal{J}}^{b}\bar{\mathcal{J}}^{c}=-\frac{i\epsilon^{bc}_{\ \ a}\bar{\mathcal{J}}^{a}}{\bar{z}-\bar{w}}
+ \frac{\frac{k}{2}\eta^{bc}}{(\bar{z}-\bar{w})^{2}}, \ee 
where $\epsilon^{abc}$ is the totally antisymmetric tensor, with
$\epsilon^{123} = 1$, and $\eta^{ab}$ is the metric defined by
$\eta^{ab}=diag(1,1,-1)$. There is a relative minus sign between the
OPEs for the left and the right moving currents, as noted
in~\cite{Martinec:2001cf}. This minus sign can be fixed using a
relabelling process, setting $J^a = \mathcal{J}^a$, $\bar J^1 = \bar{\mathcal{J}}^1$, $\bar J^3 = \bar{\mathcal{J}}^3$, $\bar J^2 = -
\bar{\mathcal{J}}^2$. The OPEs for both the left and right moving sectors are
then identical for the new currents,
\be \label{OPE}
J^{b}J^{c}=\frac{i\epsilon^{bc}_{\ \ a}J^{a}}{z-w} + \frac{\frac{k}{2}\eta^{bc}}{(z-w)^{2}}, \qquad
\bar{J}^{b}\bar{J}^{c}=\frac{i\epsilon^{bc}_{\ \ a}\bar{J}^{a}}{\bar{z}-\bar{w}} + \frac{\frac{k}{2}\eta^{bc}}{(\bar{z}-\bar{w})^{2}}.
\ee
Assuming the currents have trivial monodromies, they will have a mode
expansion 
\be
J^a=\sum_{n\epsilon\mathbb{Z}}z^{-n-1} J^a_{n} \qquad
\bar{J}^a=\sum_{n\epsilon\mathbb{Z}}\bar{z}^{-n-1}\bar{J}^a_{n}.
\ee
The commutation relations for these modes are then 
\begin{equation}
[J^a_n, J^b_m] = i \epsilon^{ab}_{\ \ c} J^c_{m+n} + \frac{k}{2} n
\eta^{ab} \delta_{m+n,0},
\end{equation}
and similarly for the $\bar{J}^a$. In particular the zero modes form
an $SL(2,\mathbb{R}) \times SL(2,\mathbb{R})$ subalgebra,
corresponding to the spacetime isometries. 

\subsection{Zero mass black hole}
\label{poin}

The $M=0$ BTZ black hole corresponds to writing the space in
Poincar\'e coordinates and making an identification.  In these
coordinates the $SL(2,\mathbb{R})$ group element is
\begin{equation}
 g = \left( \begin{array}{cc}
             \frac{1}{z} & \frac{(t+x)}{z} \\
	      \frac{(x-t)}{z} & \frac{(x^{2} + z^{2} - t^{2})}{z}
            \end{array} \right),
\end{equation}
so the metric is 
\be \label{poinmet}
ds^2 = \frac{k}{z^2}(-dt^2 +dz^2 + dx^2)
\ee
and the NSNS 2-form field is
\be \label{Bfield}
B = \frac{k}{z^2} dt \wedge dx.
\ee
In Poincar\'e coordinates, the currents are
\ba\label{Currents1}
J^1&=&-ik\left[(\partial x +\partial t)\frac{(x-t)}{z^2}+\frac{\pd z}{z}\right], \\
J^2&=& ik\left[-\frac{(x-t)}{z}\pd z +\frac{(\pd x - \pd t)}{2} + \frac{(\pd x + \pd t)}{2z^2}(2tx +1-x^2-t^2)\right],\\
J^3&=& -ik \left[-\frac{(x-t)}{z}\pd z +\frac{(\pd x - \pd t)}{2} + \frac{(\pd x + \pd t)}{2z^2}(2tx -1-x^2-t^2)\right],
\ea
and 
\ba\label{Currents2}
\bar J^1&=&-ik\left[(\bar \partial x -\bar \partial t)\frac{(x+t)}{z^2}+\frac{\bar \pd z}{z}\right], \\
\bar J^2&=& -ik\left[-\frac{(x+t)}{z}\bar \pd z +\frac{(\bar \pd x + \bar \pd t)}{2} + \frac{(\bar \pd x - \bar \pd t)}{2z^2}(-2tx +1-x^2-t^2)\right],\\
\bar J^3&=& ik \left[-\frac{(x+t)}{z}\bar \pd z +\frac{(\bar \pd x + \bar \pd t)}{2} + \frac{(\bar \pd x - \bar \pd t)}{2z^2}(-2tx -1-x^2-t^2)\right].
\ea

To relate the spacetime energy and momentum to these currents,
we consider infinitesimal time and space translations. For the
time-translation, the infinitesimal transformation is
\begin{equation} \label{deltag} 
 \delta_{(t)}g \equiv i\epsilon_{(t)}(z)g -
 ig\bar{\epsilon}_{(t)}(\bar{z}) = \frac{\partial g}{\partial t}
 \delta t, 
\end{equation}
where
\begin{equation}
 i\epsilon_{(t)}(z)= \left( \begin{array}{cc}
                            0 & 0 \\
                            -1 & 0
                           \end{array} \right) \delta t, \qquad
i\bar{\epsilon}_{(t)}(\bar{z}) = \left( \begin{array}{cc}
                                 0 & -1 \\
                                 0 & 0
                           \end{array} \right) \delta t.
\end{equation}
Thus, in terms of the $SL(2,\mathbb{R})$ generators \eqref{slgens},
\begin{equation}
\epsilon_{(t)}(z)= (\tau^{2} + \tau^{3})\delta t \equiv \tau^{+}
\delta t, \qquad
\bar{\epsilon}_{(t)}(\bar{z}) = (\tau^{2} - \tau^{3})\delta t \equiv -
\tau^{-} \delta t.
\end{equation}
A similar calculation can be performed for the infinitesimal
transformations in the $x$-direction, $\delta_{(x)}g$. Using the Ward
identity \eqref{Ward} and substituting in $\epsilon$ and
$\bar{\epsilon}$, it can then be seen that
\be \label{charges}
Q_{t}= (\mathcal{J}^{+}_{0} + \bar{\mathcal{J}}^{-}_{0}), \qquad
Q_{x}= (-\mathcal{J}^{+}_{0} + \bar{\mathcal{J}}^{-}_{0}).
\ee
In terms of the modified currents, the charges are then
\be
Q_{t}= (J^{+}_{0} + \bar{J}^{+}_{0}), \qquad
Q_{x}= -(J^{+}_{0} - \bar{J}^{+}_{0}).
\ee

We obtain the $M=0$ BTZ black hole by making periodic identifications
along the $\pd_x$ direction. The period of the identification can be
changed by rescaling $x$, so it is not a physical parameter. For
convenience, we choose $x \sim x + 2\pi$. Invariance under this
orbifold restricts states to have a quantised value of $Q_x$,
\be \label{quant}
(J^{+}_{0} - \bar{J}^{+}_{0})  \in \mathbb{Z}.
\ee
We therefore need to work in a parabolic basis for $SL(2,\mathbb{R})$,
which diagonalises $J^+_0$. In the next section, we will use a
Wakimoto representation for these currents to implement this
constraint.

\section{Vertex operators on zero mass black hole}

In this section, we construct vertex operators for the untwisted and
twisted sector states of the bosonic string on the $M=0$ BTZ black
hole. 

To implement the constraint \eqref{quant}, it is crucial to have a set
of vertex operators which diagonalise the action of $J^+_0$. It is
therefore useful to have a representation of the current algebra where
$J^+$ is as simple as possible. In the case of non-extreme BTZ black
hole in~\cite{Rangamani:2007fz}, we needed a representation which
diagonalised $J^2$, and we could simply introduce a free boson
representing the current $J^2$, writing the remainder of the vertex
operator in terms of a parafermion. Since the current $J^+$ is null, a
simple free boson representation will not be possible. However, it
turns out that the hyperbolic Wakimoto representation of the
$\widehat{SL(2,\mathbb{R})}$ current algebra introduced in~\cite{Satoh:1997xe}
provides a simple representation for $J^+$ (the relevance of this
representation for the $M=0$ BTZ black hole was previously noted
in~\cite{Troost:2002wk}).  The Wakimoto representation constructs the
conserved currents in terms of a free boson $\phi$ and anticommuting
$\beta - \gamma$ bosonic ghosts:
\ba \label{J's}
iJ^{+}(z) &=& \beta(z), \\
iJ^{-}(z) &=& \gamma^{2}\beta+\sqrt{2k'}\gamma\partial\phi(z) +
k\partial\gamma(z),\\ 
iJ^{1}(z) &=& -\gamma\beta(z) - \sqrt{\frac{k'}{2}}\partial\phi(z),
\ea
where $k'\equiv k - 2$, and the OPEs for $\beta$,$\gamma$ and $\phi$ are
\ba \label{WOPE}
\beta(z)\gamma(w) &=& -\gamma(z)\beta(w) \sim \frac{1}{z-w}, \\
\phi(z)\phi(w) &\sim& - \ln{(z-w)}.
\ea
This leads to the required OPEs for the conserved currents,
\ba
J^{+}(z)J^{-}(w) &\sim& \frac{-k}{(z-w)^{2}} + \frac{2iJ^{1}(w)}{z-w} \\
J^{1}(z)J^{\pm}(w) &\sim& \frac{\mp iJ^{\pm}(w)}{z-w} \\
J^{1}(z)J^{1}(w) &\sim& \frac{\frac{k}{2}}{(z-w)^{2}}
\ea
Unlike in the non-zero mass BTZ black hole case, this is an explicit
representation of the full current algebra. We introduce an identical
representation for the antiholomorphic currents $\bar J^a$, in terms
of $\bar \beta(\bar z)$, $\bar \gamma(\bar z)$, $\bar \phi (\bar z)$. 

\subsection{Untwisted sector vertex operators}

We want to take a basis of vertex operators which diagonalise
$J^+_0$. A vertex operator $V$ has $J^+_0$ eigenvalue $\lambda$ if
$J^{+}V(z) = \frac{\lambda V(z)}{z-w}$. Using (\ref{WOPE}), this
implies that 
\be\label{lambda}
V(z)=e^{i \lambda \gamma} f(\beta,\phi).
\ee
For AdS$_3$, the $\widehat{SL(2,\mathbb{R})}$ current algebra is a
spectrum generating algebra, and the spectrum contains short string
states in highest weight representations of the current algebra: the
continuous representations $\hat{\mathcal{C}}_j^\alpha \times
\hat{\mathcal{C}}_j^\alpha$ for $j = \frac{1}{2} + i s$ which
correspond to spacetime tachyons, and the discrete representations
$\hat{\mathcal{D}}^\pm_j \times \hat{\mathcal{D}}^\pm_j$ for
$\frac{1}{2} < j < \frac{k-1}{2}$. The spectrum in global AdS$_3$ also
contains long string states, but these do not survive the orbifold
projection, being replaced instead by the twisted sector
states. Diagonalising $J^+_0$ corresponds to considering the
representations of $SL(2,\mathbb{R})$ in a parabolic basis. For both
the continuous and discrete representations of the current algebra, in
this parabolic representation, the eigenvalue $\lambda$ can take all
real values.

Since these are highest weight representations of the current algebra,
we can focus on the chiral primary operators; other vertex operators
will be obtained as descendents. Requiring that \eqref{lambda} be a
chiral primary operator implies that $f(\beta,\phi)$ is independent of
$\beta$, as $V$ would otherwise have too singular an OPE with
$J^-$. Including the anti-holomorphic sector, we can therefore take a
basis of chiral primary vertex operators in the untwisted sector of
the form
\be \label{unvo}
V_{j \lambda \bar \lambda}(z, \bar z) = e^{i \lambda\gamma
  -\sqrt{\frac{2}{k'}}j\phi} e^{i \bar \lambda \bar \gamma
  -\sqrt{\frac{2}{k'}}j \bar \phi} .
\ee
To implement the orbifold, we need to
quantise the eigenvalue of $J^+_0 - \bar J^+_0$, that is, we need
\begin{equation} \label{chquant}
\lambda - \bar{\lambda} \in \mathbb{Z}. 
\end{equation}
In the next subsection, we will see how this quantisation condition
can be implemented using a twist operator. This will also allow us to
construct vertex operators for the twisted sector modes.

The energy momentum tensor for the WZW model is 
\be
T=\frac{1}{k-2}\eta_{ab}:J^{a}J^{b}:,
\ee
which can be rewritten in in terms of $J^+$ and $J^-$ as
\be\label{T}
T = \frac{1}{(k-2)}:\left(J^{1}J^{1}-\frac{1}{2}J^{+}J^{-}
-\frac{1}{2}J^{-}J^{+}\right):\ .
\ee
Working in the Wakimoto representation,  
\be\label{T2}
T= \beta\partial\gamma -\frac{\partial^{2}\phi}{\sqrt{2k'}} -
\frac{(\partial\phi)^{2}}{2}.
\ee
The conformal dimensions of the vertex operators \eqref{unvo} are then 
\be
h= \bar h = \frac{-j(j-1)}{(k-2)}.
\ee
Thus, the label $j$ on the vertex operators corresponds to the label
on representations of the current algebra.

\subsection{Twist operator}

So far, we have just described the vertex operators describing strings
on AdS$_3$ in a basis which is adapted to working in Poincar\'e
coordinates. To describe the $M=0$ BTZ black hole, we would now like
to impose the quantisation condition \eqref{quant}. Following the same
route as in the analysis of the non-extremal BTZ black
hole~\cite{Rangamani:2007fz}, we would like to impose this condition
by requiring mutual locality of the untwisted sector vertex operators
\eqref{unvo} with an appropriate twist operator. The twisted sector
vertex operators will then be obtained by closure of the OPE including
the twist operator. 

To do this we have to change our representation again and bosonize the
$\beta-\gamma$ system as in~\cite{Satoh:1997xe},
\ba
\beta &=& \partial\phi_{+}, \\
\gamma &=& \phi_{-},
\ea
where
\be
\phi_{\pm} = \frac{1}{\sqrt{2}} (\phi_{0} \pm \phi_{1}), 
\ee
\be
\phi_{i}(z)\phi_{j}(w) \sim -\eta_{ij}\ln(z-w), \qquad \eta_{ij} =
diag(-1,1). 
\ee
We introduce a similar bosonization for $\bar \beta, \bar \gamma$. In
this representation, the untwisted sector chiral primary fields become
\be
V_{j \lambda \bar \lambda}(z, \bar z) = e^{i\lambda\phi_{-}
  -\sqrt{\frac{2}{k'}}j\phi} e^{i\bar \lambda \bar \phi_{-}
  -\sqrt{\frac{2}{k'}}j\bar \phi}.
\ee
This representation was used to discuss strings on AdS$_3$ in
Poincar\'e coordinates in~\cite{Bars:1995mf}. It was noted there that
there are potential logarithmic branch cuts associated with the
definition of the boson $\phi_+$. We will now see (as also noted
in~\cite{Troost:2002wk}) that these branch cuts are correctly interpreted as
reflecting winding around the $x$ direction, describing the twisted
sectors in the string on the $M=0$ BTZ black hole.  

The appropriate twist operators are
\be \label{twist}
t_{n} = e^{in(\phi_{+}+\bar \phi_+)}.
\ee
This will impose the correct quantisation condition, as
\be
t_{n}(z)V_{j \lambda \bar \lambda}(w) \sim
\frac{\exp(in\phi_{+} + i\lambda\phi_{-}
  -\sqrt{\frac{2}{k'}}j\phi)}{(z-w)^{n\lambda}}
\frac{\exp(in \bar \phi_{+} + i\bar
    \lambda \bar \phi_{-} -\sqrt{\frac{2}{k'}}j \bar
  \phi)}{(\bar z- \bar w)^{n \bar \lambda}}, 
\ee
so the OPE will only be mutually local for $\lambda - \bar \lambda \in
\mathbb{Z}$. 

We can also read off the twisted sector vertex operators from this
OPE, obtaining 
\be
V_{j n \lambda \bar \lambda } = \exp(in\phi_{+} +
i\lambda\phi_{-} -\sqrt{\frac{2}{k'}}j\phi)
\exp(in\bar \phi_{+} + i\bar
  \lambda\bar \phi_{-} -\sqrt{\frac{2}{k'}}j \bar \phi).
\ee
Note that the current algebra generators $J^-, J^2$ will have
non-trivial monodromies around a twisted sector vertex operator
because of the dependence on $\phi_+$, reflecting the twisting.  The
conformal dimensions for these twisted sector operators are
\be
h=- \frac{j(j-1)}{k'} -n \lambda, \quad \bar h = - \frac{j(j-1)}{k'} -
n \bar \lambda.
\ee
The level-matching condition $h - \bar h \in \mathbb{Z}$ is satisfied
as a consequence of the quantisation of the charge $Q_x$ in
\eqref{chquant}. Comparing this to the spectrum obtained for the
twisted sectors of the non-extreme BTZ black hole
in~\cite{Rangamani:2007fz}, we see that the spectrum there reduces to
this one in the limit $r_+ \to 0$, as expected, with $r_+
\lambda_{there} = - \lambda_{here}$, $r_+ \bar \lambda_{there} = \bar
\lambda_{here}$.  The full vertex operators involve taking descendents
of these chiral primary operators, and include a contribution from the
internal CFT. The physical state conditions will then be
\be
- \frac{j(j-1)}{k'} -n \lambda + h_{int} + N = - \frac{j(j-1)}{k'} -
n \bar \lambda + \bar h_{int} + \bar N =1, 
\ee
where $h_{int}, \bar h_{int}$ are the dimensions of the operator from
the internal CFT, and $N, \bar N$ are oscillator numbers for the
current algebra. We assume that the internal CFT is unitary, so
$h_{int}, \bar h_{int} \geq 0$.

This is one of the main results of our paper: by adopting this
bosonised version of the Wakimoto representation for the currents, we
see that we can give completely explicit expressions for the vertex
operators for the full spectrum of string states on the M=0 BTZ black
hole, including twisted sector states. In \cite{Troost:2002wk}, this
representation of the currents was also used to construct the Virasoro
generators associated with the asymptotic isometries of the spacetime
in terms of the worldsheet currents. This makes it possible to study
the relation of the worldsheet theory to the dual CFT in AdS/CFT, as
was done for global AdS$_3$ in \cite{Giveon:1998ns,deBoer:1998pp}.

\subsection{Tachyons in twisted sectors} \label{tachyon}

As we are working with the bosonic string, we know there is a
tachyonic ground state in the untwisted sectors. We would like to see
if there is a tachyon in the twisted sectors. In the extension to the
supersymmetric case, the expectation is that there will be a choice of
GSO projection which eliminates the ground state in the untwisted
sectors but retains it in the odd twisted sectors (corresponding to
choosing an antiperiodic spin structure on the orbifold circle in
spacetime). Given the appearance of a tachyon in twisted sectors for
the non-extreme BTZ black hole with $\sqrt{k} r_+
<2$~\cite{Rangamani:2007fz}, we would expect that there will be one
here as well.

We need to consider carefully the definition of a
tachyon. Classically, a spacetime field is tachyonic if it has
normalisable solutions which grow exponentially in time. It is
difficult to look directly for modes which grow exponentially in time
from the worldsheet point of view, as this would require complex
$\lambda, \bar \lambda$, which makes it difficult to see how we can
satisfy the physical state condition $h = \bar h= 1$ in the twisted
sectors. We will therefore look instead for modes with zero energy; if
a spacetime field has a normalisable solution of zero energy, it
should generically also have solutions which grow exponentially in
time, by continuity.\footnote{In a black hole background, modes
  supported close to the horizon have low energy as a result of the
  gravitational redshift, so a non-tachyonic field will have an energy
  spectrum starting from zero. However, it will not have a
  normalisable mode of strictly zero energy, so we believe this
  condition is still physically appropriate even in the presence of a
  black hole horizon.} Since the $x$ direction is spacelike everywhere
in the $M=0$ BTZ black hole, it is physically reasonable to further
restrict to modes which also have zero momentum along $x$; these
should be the most tachyonic modes for a field with a given
mass-squared. Thus, it suffices for us to consider modes with $\lambda
= \bar \lambda =0$.

Thus, we are looking for normalisable modes with $\lambda = \bar
\lambda =0$, satisfying the physical state condition. In this case,
the physical state condition for the twisted sectors is identical to
that in the corresponding untwisted sector, and the twisted sector
states will satisfy the physical state condition whenever the
corresponding untwisted sector states do. In particular, there are
physical states obtained by spectral flow from the tachyon in the
untwisted sector, which have $j(j-1) <0$.

The remaining condition is normalisability. In the untwisted sector,
we know that we have the usual bosonic string tachyon, which has
normalisable solutions with $\lambda = \bar \lambda =0$. The twist
operator is expressed in terms of the Wakimoto representation, and not
in terms of the spacetime coordinates, so it is not possible to
rigorously relate the normalisability of twisted sector modes to the
corresponding untwisted sector ones, but we expect that at least for
large $k$, the twisting will not significantly modify the dependence
on the radial coordinate, so twisted sector modes will be normalisable
if the corresponding untwisted sector mode is. Thus, we expect that
the $\lambda = \bar \lambda =0$ vertex operators in twisted sectors
obtained from the tachyon in the untwisted sector will be
normalisable, and hence give modes of a tachyon in the twisted
sectors. 

As in \cite{Rangamani:2007fz}, these modes have roughly the
same radial behaviour as for the untwisted sector tachyon, so they are
not well-localised in the neighbourhood of the horizon. The twisted
sector states are essentially long string states, which can propagate
to the asymptotic boundary of the $M=0$ BTZ black hole at low cost in
energy because of the coupling the NSNS 2-form field. 

The expectation is that while the untwisted sector tachyons are
removed by the GSO projection in the supersymmetric theory, the
untwisted sector tachyons will remain. We will comment on this again
when we discuss the extension of our work to the superstring later.

\section{The Null orbifold limit}

An important source of intuition and a useful check on the
calculations in studying orbifolds of AdS$_3$ is to consider the limit
$k \to \infty$, in which the space becomes flat. For the non-zero mass
BTZ black hole studied in~\cite{Rangamani:2007fz}, there were two flat
space limits of interest, the near horizon limit which focused on the
region near the event horizon, and the Milne limit, which focused on
the singularity. For the $M=0$ BTZ black hole the event horizon and
the singularity are at the same point in space, $z=\infty$ in the
coordinates of \eqref{poinmet}. The two limits are therefore replaced
by one, the null orbifold limit. In this section, we consider the
behaviour of the untwisted and twisted sector states we constructed
above as we take this limit. This limit is most closely analogous to
the Milne limit in~\cite{Rangamani:2007fz}.

To show that the $M=0$ BTZ black hole reduces to the null orbifold as
we take $k \to \infty$ focusing on the region near the Poincar\'e
horizon at $z=\infty$, we need to make a change of coordinates. If we
define new coordinates
\be \label{ycoords}
y^+ = \frac{\sqrt{k}}{z}, \quad y^- = \sqrt{k} (t+z), \quad y = x, 
\ee
then the metric \eqref{poinmet} becomes
\be \label{nullomet}
ds^{2}= -\frac{(y^+)^2 (dy^-)^2}{k} - 2dy^+ dy^- + (y^+)^2 dy^2.
\ee
Taking the limit $k \to \infty$ for fixed $y^\pm, y$, the first term
vanishes, and we can see that the metric reduces to the null orbifold
of~\cite{Steif:1995zm,Simon:2002ma}, which was analysed in string
theory in~\cite{Liu:2002ft,Liu:2002kb}. In these coordinates, the null
orbifold is simply the identification $y \sim y + 2 \pi$. The
quantization condition associated with this orbifold identification
remains simply $\lambda - \bar \lambda \in \mathbb{Z}$. We should also
consider the limit for the NSNS 2-form field 
\eqref{Bfield}. To obtain a finite limit as $k \to \infty$, we first
need to make a gauge transformation to write
\begin{equation}
B = \frac{k}{z^2} (dt+dz) \wedge dx = \frac{(y^+)^2}{\sqrt{k}} dy^- \wedge dy, 
\end{equation}
so in this gauge the 2-form vanishes in the limit as $k \to
\infty$. The contribution from the 2-form is still important to see
that the currents $J^a$, $\bar J^a$ are conserved to sub-leading order
as we take the limit, as in~\cite{Rangamani:2007fz}. 

It is also
convenient to rewrite the null orbifold in Cartesian coordinates; the
relation is
\begin{equation}
x^+ = y^+, \quad x^- = y^- + \frac{1}{2} y^+ y^2, \quad x = y^+ y.
\end{equation}
In these coordinates, the null orbifold metric is simply flat, but the
identification is more complicated: 
\begin{equation} \label{xorb}
(x^+, x^-, x) \sim (x^+, x^- + 2\pi x + 2 \pi^2 x^+, x + 2\pi x^+). 
\end{equation}

Using \eqref{Currents1} the currents can be calculated in terms of
$y^\pm, y$: 
\ba
J^1&=&-i\sqrt{k}(y^+ \pd y + y \pd y^+) -iy^+ \pd y^- +i y^- \pd y^+
-i (y^+)^2 y \pd y +O(\frac{1}{\sqrt{k}}),\\
J^2&=&-i\sqrt{k} (y^+ y \pd y +\pd y^-
+\frac{y^2 \pd y^+}{2} -\frac{\pd y^+}{2}) \\ 
& & + i(-\frac{(y^+)^2 y^2 \pd y }{2} + y^+ y^- \pd y + \frac{(y^+)^2
  \pd y}{2} - y^+ y \pd y^-  + y^- y \pd y^+ )
+O(\frac{1}{\sqrt{k}}), \nonumber \\ 
J^3&=&i\sqrt{k} (y^+ y \pd y +\pd y^-
+\frac{y^2 \pd y^+}{2} +\frac{\pd y^+}{2}) \\ 
& & -i(-\frac{(y^+)^2 y^2 \pd y }{2} + y^+ y^- \pd y - \frac{(y^+)^2
  \pd y}{2} - y^+ y \pd y^-  + y^- y \pd y^+ )
+O(\frac{1}{\sqrt{k}}). \nonumber
\ea
This can be more simply re-expressed in terms of $x^\pm, x$:
\ba
J^1&=&-i\sqrt{k}\pd x -ix^+ \pd x^- +i x^- \pd x^+ +O(\frac{1}{\sqrt{k}}),\\
J^2&=&-i\sqrt{k} (\pd x^- -\frac{\pd x^+}{2})  + i(x^- \pd x - x \pd
x^- + \frac{1}{2} (x^+ \pd x - x \pd x^+) )
+O(\frac{1}{\sqrt{k}}),\\ 
J^3&=&i\sqrt{k} (\pd x^- +\frac{\pd x^+}{2})  - i(x^- \pd x - x \pd
x^- - \frac{1}{2} (x^+ \pd x - x \pd x^+) )+O(\frac{1}{\sqrt{k}}).
\ea
We see that in the flat space limit $k \to \infty$, the leading order
($O(\sqrt{k})$) terms reproduce the Cartesian translation currents on
flat space. As in~\cite{Rangamani:2007fz}, the subleading parts
involve Lorentz transformations in the flat space limit, and are
required to make the total current conserved to subleading order
taking into account the effects of the 2-form field. The same can
be done for the antiholomorphic sector.

We can use this expression for the currents to relate the Wakimoto
variables to the coordinates in this limit, 
\ba
\beta &=& -\sqrt{k} \pd x^+ - (x^+ \pd x - x \pd x^+), \\
\gamma &=& \frac{2}{\sqrt{k}}x^- +\frac{2}{k}x x^- ,\\ 
\phi &=& -\sqrt{\frac{2}{k'}}\left(\sqrt{k} x  + x^+ x^- \right). 
\ea
This in turn implies that the bosonised versions of the Wakimoto
variables are closely related to the $x^\pm, x$ coordinates in the
flat space limit; to leading order, 
\begin{equation} \label{waklimit1}
\phi^+ = - \sqrt{ k} x^+, \quad \phi^- = \frac{2}{\sqrt{k}}
  x^-, \quad \phi = - \sqrt{2} x. 
\end{equation}
By studying the flat space limit of the antiholomorphic currents $\bar
J^a$, one can similarly learn that
\begin{equation}\label{waklimit2}
\bar \phi^+ = - \sqrt{ k} \bar x^+, \quad \bar \phi^- = - \frac{2}{\sqrt{k}}
  \bar x^-, \quad \bar \phi = \sqrt{2} \bar x
\end{equation}
to leading order. Note that the factors of $2$ in these expressions
appear because in units with $\alpha' =1$, the flat space coordinates
have OPEs $x^\mu x^\nu \sim \frac{1}{2} \eta^{\mu\nu} \ln (z-w)$. 

\subsection{States and vertex operators}

In the untwisted sector, the states which survive in this flat space
limit are those with $\lambda - \bar \lambda \in \mathbb{Z} \sim
\mathcal{O}(1)$ and $\lambda + \bar \lambda \sim
\mathcal{O}(\sqrt{k})$ (since we hold $y^- = \sqrt{k}(t+r)$ fixed in
the limit). To have $h \sim \mathcal{O}(1)$, we need $j \sim
\mathcal{O}(\sqrt{k})$, and similarly for the barred quantities. In
the twisted sectors, we still have $\lambda - \bar \lambda \in
\mathbb{Z} \sim \mathcal{O}(1)$ and $\lambda + \bar \lambda \sim
\mathcal{O}(\sqrt{k})$, since the $Q_x$ and $Q_t$ eigenvalues are
unaffected by twisting. However, in twisted sectors $h =
-\frac{j(j-1)}{k-2} - n \lambda$, so $h \sim \mathcal{O}(1)$ in
twisted sectors requires $j \sim \mathcal{O}(k^{3/4})$ to cancel the
$\sqrt{k}$ contribution from $\lambda$. This cancellation can be
achieved for modes in the continuous representations of the current
algebra if $\lambda >0$, and for modes in the discrete representations
of the current algebra if $\lambda <0$. Thus, both tachyonic and
non-tachyonic twisted sector modes survive in the flat space limit,
but with this curious correlation with the sign of $\lambda$. We
expect that the resulting spectrum in the flat space limit should
agree with the one obtained in~\cite{Liu:2002ft}.

We note that as in the Milne limit in~\cite{Rangamani:2007fz}, in
general the modes which survive in twisted sectors in this flat space
limit are not the ones which are obtained by twisting from the
untwisted sector states which survive in the limit. This can also be
seen by observing that the twist operator \eqref{twist} becomes, to
leading order,
\begin{equation}
t_{n} = e^{-n \sqrt{k}(x_{+}+\bar x_+)},
\end{equation}
and hence does not have a well-defined flat space limit. Thus, our
twist operator construction does not have a counterpart in the null
orbifold. 

Despite the failure of the twist operator to survive in the flat space
limit, one might still hope that we could follow our vertex operators
in this limit, since we have an explicit construction of the vertex
operators in terms of the Wakimoto representation and we understand
how these Wakimoto fields are related to flat space coordinates in the
limit. Disappointingly, this does not work. If we consider the
vertex operator \eqref{unvo} and substitute in the leading order
relations between the Wakimoto fields and the coordinates
\eqref{waklimit1}, we obtain 
\be
V_{j \lambda\bar{\lambda}}(z) = e^{i\frac{2\lambda}{\sqrt{k}}x^{-} +
  j\frac{2\sqrt{k}}{k-2}x}e^{-i\frac{2\bar{\lambda}}{\sqrt{k}}\bar{x}^{-}
  - j\frac{2\sqrt{k}}{k-2}\bar{x}}. 
\ee
In the limit, let us write $2\lambda = \sqrt{k} E + P_x$,
$2\bar{\lambda} = \sqrt{k} E - P_x$, $j = \sqrt{k} m$. Then the vertex
operator is to leading order 
\begin{equation} \label{limvo} 
V_{j \lambda\bar{\lambda}}(z) = e^{iE x^{-} +
  2 m x}e^{-i E \bar{x}^{-}
  - 2 m\bar{x}}.  
\end{equation}
We see that this expression has lost the dependence on $P_x$, so it
degenerates in the flat space limit. This failure to obtain a good
representation of the vertex operators in the flat space limit is in
retrospect not unexpected: the $\widehat{SL(2,\mathbb{R})} \times
\widehat{SL(2,\mathbb{R})}$ structure we used in constructing our
vertex operators degenerates in this limit. Although the metric and
spectrum smoothly go over to the flat space orbifold in this limit, a
new representation of the vertex operators is necessary. Similarly,
when we take the $r_+ \to 0$ limit of the non-zero mass BTZ black
hole, the spectrum of~\cite{Rangamani:2007fz} reduces to the one we
have obtained here for the zero-mass black hole, but the vertex
operators do not have a smooth limit. A new representation is needed
in the limit, reflecting the fact that we are considering a different
orbifold.   

Thus, while the orbifold we are considering reduces to the null
orbifold in the limit $k \to \infty$, the representation of the vertex
operators in terms of Wakimoto fields degenerates in this limit, so we
do not seem to be able to glean new insight into the null orbifold
from our construction. 

\section{The extremal rotating black hole}

We can easily extend this investigation of the zero mass BTZ black
hole to study extremal rotating BTZ black holes, with $M=J$. Rotating
BTZ black holes are orbifolds of AdS$_3$ with an asymmetric action on
the worldsheet; as we will see below, the extremal rotating black
holes correspond to an orbifold where the action on the left-movers is
the same as for the zero mass BTZ black hole, while the action on the
right movers is the same as for the non-zero mass black hole studied
in ~\cite{Rangamani:2007fz}. Thus, by combining our previous results,
it is easy to determine the spectrum in this case as well. The
construction of the twisted sectors is based on an appropriate ansatz,
for which we then explicitly check the mutual locality.

The extremal rotating BTZ black hole has the metric
\be \label{erotmet}
ds^2 = k\left(-\frac{(r^2-r_+^2)^2}{r^2}d\tau^2 +
  \frac{r^2}{(r^2-r_+^2)^2}dr^2 + r^2(d\phi -
  \frac{r_+^2}{r^2}dt)^2\right),  
\ee 
where $\phi$ is a periodic coordinate, $\phi \sim \phi +2\pi$. The
spacetime is locally AdS$_3$, with periodicity of $\phi$ corresponding
to the action of an orbifold. If we write AdS$_3$ in the embedding
coordinates $x^0, x^1, x^2, x^3$ in $\mathbb{R}^{2,2}$ as the
hyperboloid $-x_0^2 - x_1^2 + x_2^2 + x_3^3 = -k^2,$ the orbifold
which gives us the above extremal rotating black hole is along the
Killing vector~\cite{Banados:1992gq} 
\be
\xi = r_+ (J_{03} + J_{12}) + J_{01} - J_{02} - J_{13} + J_{23},
\ee
up to conjugation, where $J_{ab}$ are the Lorentz transformations on
$\mathbb{R}^{2,2}$, $J_{ab} = \eta_{bc} x^c \pd_a - \eta_{ac} x^c
\pd_b$. In the coordinate system of \eqref{erotmet}, this Killing
vector is $\xi = \partial_\phi$. 

The extremal rotating BTZ metric is locally AdS$_3$, so it can be
related to the Poincar\'e coordinate system we used earlier. The
coordinate transformation
\begin{equation}
z = R^{-1/2} e^{r_+ (\phi-t)}, \quad t+x = e^{2 r_+ (\phi-t)}, \quad t-x =
- (T+\phi), 
\end{equation}
where
\be
R=\frac{1}{2r_+}(r^2-r_+^2), \quad 
T=t - \frac{r_+}{r^2-r_+^2}, 
\ee
converts the metric \eqref{poinmet} into the metric
\eqref{erotmet}. This corresponds to writing the group element of
$SL(2,\mathbb{R})$ as
\be
g = \left( \begin{array}{cc}
             R^{\frac{1}{2}}e^{-r_+(\phi-t)} &  R^{\frac{1}{2}}e^{r_+(\phi-t)}  \\
	     (T + \phi)R^{\frac{1}{2}}e^{-r_+(\phi-t)} & R^{\frac{1}{2}}e^{r_+(\phi-t)}\left(T + \phi + \frac{1}{R}\right)
            \end{array} \right).
\ee
As in section \ref{poin}, we can determine the conserved charge
associated with $\phi$ translation. The action of $\phi$ translation
on the group element is 
\begin{equation}
\delta_{(\phi)} g = \left( \begin{array}{cc} 0 & 0 \\ 1 &
    0 \end{array} \right) g + r_+ g \left( \begin{array}{cc} -1 & 0 \\ 0 &
    1 \end{array} \right),
\end{equation}
so the conserved charge is 
\be \label{qphirot}
Q_{\phi} = -J_0^+ + r_+ \bar{J}_0^1. 
\ee
Note that this naive expression will apply for the untwisted sectors;
for the twisted sectors, there is the possibility of a total
derivative term, which we need to determine.

Thus, we see that the action in terms of $SL(2,\mathbb{R}) \times
SL(2, \mathbb{R})$ is chiral, with the left moving part looking like
that of a massless BTZ black hole we have studied above, while the
right moving part looks like that of the massive BTZ black hole. It is
therefore natural to choose the parafermionic representation for the
right movers, and the Wakimoto representation introduced above for the
left movers. The parafermionic representation for the right movers
involves writing the currents as
\begin{equation}
\bar J^1 = -i \sqrt{\frac{k}{2}} \pd \bar X, \quad \bar J^\pm = \bar
\xi^\pm e^{\pm   \sqrt{\frac{2}{k}} \bar X},    
\end{equation}
where $\bar X$ is a free boson, $\bar X(\bar z) \bar X(\bar w) \sim -
\ln(\bar z-\bar w)$, and $\bar \xi^\pm$
are parafermions representing the remaining
$\widehat{SL(2,\mathbb{R})}_k/\widehat{U(1)}$ algebra.  Thus, the
vertex operators in the untwisted sector are
\be
V_{j \lambda \bar \lambda}(z) = e^{i\lambda\gamma -j\sqrt{\frac{2}{k'}}\phi}\bar{\Psi}_{j\bar{\lambda}}e^{-i\sqrt{\frac{2}{k}}\bar{\lambda}\bar{X}}
\ee
Where $\bar{\Psi}_{j\bar{\lambda}}$ are parafermionic operators with
conformal dimension $\bar h_{\bar \Psi} = -\frac{j(j-1)}{k'} -
\frac{\bar \lambda^2}{k}$. We know that these untwisted sector vertex
operators are mutually local for the operators corresponding to modes
of fields on AdS. This gives us some information about the OPE of the
parafermionic operators, as the OPE of two such vertex operators will
only be mutually local if 
\be \label{untwist}
m-\frac{2}{k'}jj' - \frac{2}{k}\bar{\lambda}\bar{\lambda'} \in \mathbb{Z}
\ee
where $m$ characterises the leading singularity in the OPE of the
parafermionic operators, 
\be 
\bar{\Psi}_{j\bar{\lambda}}\bar{\Psi'}_{j'\bar{\lambda'}} \sim
\frac{O}{(z-w)^m}. 
\ee

We construct an ansatz for the twisted sector states in this orbifold
by combining the results for the twisted sectors from our earlier
analysis of the massive and massless black holes: that is, we guess
that the twisted sector vertex operators are simply the right moving
part of the twisted sector state from the massive black hole,
combined with the left moving part of the twisted sector state from
the massless black hole. This gives
\be \label{tspec}
V_{j\lambda\bar{\lambda}n}(z)= \exp\left(in\phi_+ + i\lambda\phi_-
  -j\sqrt{\frac{2}{k'}}\phi\right)\bar\Psi_{j\bar{\lambda}}\exp\left(-i\sqrt{\frac{k}{2}}\left[\bar{\lambda}
    - \frac{k}{2} n r_+\right]\bar{X}\right)
\ee
The conformal dimensions for this operator are
\begin{equation}
h = - \frac{j(j-1)}{k'} -n \lambda, \quad \bar h = -\frac{j(j-1)}{k'}
- \bar \lambda r_+ n + \frac{kn^2 r_+^2}{4}.  
\end{equation}
As in section \ref{tachyon}, we should have a tachyon in the
twisted sectors if we can satisfy the physical state condition for a
mode with $j(j-1) <0$ and $\lambda = \bar \lambda =0$. This requires
$\sqrt{k} r_+ < 2$, as in~\cite{Rangamani:2007fz}, so there is a
tachyon if the size of the spatial circle at the black hole horizon is
small enough. As in section \ref{tachyon}, this tachyon will not be well
localised in the region near the horizon.

Level matching will require that $h-\bar{h} \in \mathbb{Z}$, which
implies
\be \label{rotquant}
-n \lambda + r_+ n \bar{\lambda} - \frac{kn^2r_+^2}{4} \in \mathbb{Z}.
\ee
We would like to see this arise as a consequence of the quantisation
of angular momentum imposed by the orbifold. Naively, the generator of
translation in $\phi$ is \eqref{qphirot}, which would imply a
quantisation condition $ -\lambda + r_+ (\bar \lambda - k n r_+/2) \in
\mathbb{Z}$, which does not agree with \eqref{rotquant}. Hence, as in
\cite{Hemming:2001we}, we will need to include a total derivative term
in the definition of $Q_\phi$, so that in twisted sectors
\begin{equation}
Q_\phi = - J_0^+ + r_+ \bar{J}_0^1 + \frac{knr_+^2}{4}, 
\end{equation}
implying a quantization condition
\begin{equation}
- \lambda + r_+ \bar{\lambda} - \frac{kn r_+^2}{4} \in \mathbb{Z},
\end{equation}
consistent with \eqref{rotquant}.  Note that unlike in non-rotating
cases, the quantization condition on $\lambda, \bar \lambda$ depends
on the twist $n$, so the twisted sector states cannot be obtained by
considering the OPE of untwisted sector states with an appropriate
twist operator. Such a procedure would get the quantization condition
wrong.

We then need to verify mutual locality of these twisted sector
states. To determine mutual locality consider an OPE between
$V_{j\lambda\bar{\lambda}n}(z)$ and
$V_{j'\lambda'\bar{\lambda'}n'}(w)$. It turns out that the
relevant condition for mutual locality is
\be
m-\frac{2}{k'}jj'-n\lambda'- n\lambda -
\frac{2}{k}\left(\bar{\lambda} -
  \frac{k}{2} n r_+\right)\left(\bar{\lambda'} - \frac{k}{2} n'r_+\right)
\in \mathbb{Z}.
\ee
This is satisfied as a consequence of \eqref{rotquant} and
\eqref{untwist}. Therefore the correct spectrum for the twisted sector
is indeed given by \eqref{tspec}; this completes the untwisted sector
spectrum to obtain a mutually local set of operators with the
appropriate set of twisted sectors, indexed by a single integer
denoting the twist. It would be interesting to see if the modular
invariance of the resulting partition function could be explicitly
verified, but this will be complicated to check, as is generally the
case for asymmetric orbifolds, so we will not attempt to do so
explicitly here.

\section{Supersymmetry}

In this section, we will discuss the extension of our analysis of the
spectrum to the superstring. This analysis is particularly interesting
for the $M=J$ extremal rotating black holes considered here, as these
are supersymmetric backgrounds for the superstring. Also, from the
point of view of considering the tachyons, the tachyons in twisted
sectors are most interesting in the superstring, where we expect the
GSO projection to eliminate the tachyon in the untwisted sector. For
the supersymmetric choice of spin structure on spacetime, spacetime
supersymmetry implies that there are no tachyons, but if we make the
opposite choice of spin structure, we expect the tachyons in odd
twisted sectors to survive the GSO projection. 

Unfortunately, we are not able to construct the spectrum for the
superstring explicitly. We have not succeeded in extending the nice
representation of $SL(2,\mathbb{R})$ in terms of Wakimoto fields to
the superstring, so we have not succeeded even in constructing an
explicit representation for the untwisted sector modes which survive
the orbifold projection condition. In this section, we will describe
the problem, focusing on the case of the $M=0$ black hole for
simplicity. 

We consider Type II String theory on BTZ $\times \mathbf{S^{3}} \times
\mathbf{T^4}$ as described by a $\widehat{SL(2,\mathbb{R})}$ super-WZW
model at level $k$. The super-current algebra is
\be \label{totcurr}
J^a = j^a - \frac{i}{k}\epsilon^a_{\ bc}\psi^b\psi^c
\ee
with $\psi^a$ having the usual OPE structure for a fermion,
\ba
\psi^a(z)\psi^b(w) &\sim& \frac{k}{2}\frac{\eta^{ab}}{(z-w)} \\
j^a(z)\psi^b(w) &\sim& 0
\ea

The OPEs for the bosonic currents $j^a$ are almost identical to the
previous section, while the OPEs for the supercurrent $J^a$ are very similar,
\ba
j^a(z)j^b(w) &\sim& \frac{\tilde{k}}{2}\frac{\eta^{ab}}{(z-w)^2} + i\frac{\epsilon^{ab}_{\ \ c}j^c}{(z-w)} \\
J^a(z)J^b(w) &\sim& \frac{{k}}{2}\frac{\eta^{ab}}{(z-w)^2} + i\frac{\epsilon^{ab}_ {\ \ c}j^c}{(z-w)}
\ea
where $\tilde{k} = k+2$. The world-sheet $\mathcal{N} = 1$ super-current is
\be \label{super-current}
G(z) = \frac{2}{k}\left(g_{ab}\psi^aj^b -
  \frac{i}{3k}\epsilon_{abc}\psi^a\psi^b\psi^c\right). 
\ee

For the superstring, the generator of $x$ translations is again
\eqref{charges}, but where $J^a$ are now the total currents given in
\eqref{totcurr}. We therefore want to find a representation of the
total currents which simplifies the action of $J^+$, and use this
representation to write the untwisted sector vertex operator in a form
which diagonalises the action of $J^+$. We can use a Wakimoto
representation as before for the bosonic currents $j^a$, but the
fermionic contribution to $J^a$ is more problematic. 

In
\cite{Martinec:2001cf,Rangamani:2007fz}, the fermions were rewritten
in terms of a set of bosons $H_I$, $I=1, \ldots, 5$, with OPEs 
\be
H_I(z)H_J(w) = -\delta_{IJ} \ln(z-w),
\ee
such that the spin fields 
\be
S_{\alpha} = e^{\frac{i}{2}\epsilon_IH_I}
\ee
with $\epsilon_I = \pm 1$ diagonalise the action of $J^2$ and hence
$Q_\phi$. In our case, it is $J^+$ which is relevant, and this involves a
combination $\psi_1 (\psi_3 + \psi_2) \equiv \psi_1 \psi_+$. We could
formally define a field $H_*$ by
\be
\partial H_* = \psi_1\psi_+ ,
\ee
but this will not have the OPE of a free boson, so we cannot use it in
constructing the spin fields in a way analogous to
\cite{Martinec:2001cf,Rangamani:2007fz}. As a result, we have no
simple route to constructing an appropriate basis of spin fields in
this case which diagonalises the action of $Q_x$. We leave this as
an open problem for future work. It is clearly very interesting to try
to understand these simple examples of BPS black holes from the
worldsheet perspective, so we hope further progress will be possible.

\section{Discussion}

The main result of this paper is that we have obtained the full
spectrum and discussed the tachyons appearing in this spectrum for the
bosonic string on the $M=0$ BTZ black hole and on the extremal
rotating $M=J$ black hole. The spectrum on the zero mass black hole is
just the limit of the spectrum obtained for the massive black hole
in~\cite{Rangamani:2007fz}, as we would expect. However, because the
zero mass black hole corresponds to a parabolic orbifold of AdS$_3$,
the description of the states in this case is quite different from
that of~\cite{Rangamani:2007fz}. The use of the Wakimoto
representation in the $M=0$ BTZ black hole enables us to give a fully
explicit description of the vertex operators for states in both
untwisted and twisted sectors. This makes this example a particularly
interesting laboratory for further explorations of worldsheet string
theory on these black hole backgrounds; compared to the parafermionic
representation of the vertex operators employed for the massive black
hole in~\cite{Rangamani:2007fz}, this more explicit representation
ought to give us greater control. Unfortunately, however, this
description of the vertex operators appears to degenerate in the flat
space limit, so our understanding of the limit where we zoom in on the
region near the singularity is not significantly improved by the use
of this representation.

For the zero mass black hole, we argued that the twisted sector modes
obtained from the bulk tachyon in the untwisted sector are also
tachyonic. The analysis of these tachyons will closely parallel the
corresponding analysis in~\cite{Rangamani:2007fz}, so we have not
given much detail in our discussion of the tachyons. These modes are
not localised in the region near the horizon, as the coupling to the
NSNS 2-form field makes a negative contribution to the energy of the
string, allowing it to propagate to large distances. The study of the
condensation of these tachyons would be an interesting direction for
further work, but because the tachyon is not localised, it may be
quite challenging. 

We extended our work on the zero mass black hole by considering the
extremal rotating black hole, which corresponds to an asymmetric orbifold,
with the action on left movers the same as for the zero mass black
hole and the action on right movers the same as for the massive
non-rotating black hole. We were therefore able to construct a
proposal for the spectrum of strings on this background by combining
our work on the zero mass black hole with previous work on the massive
black hole. 

Finally, we considered the extension to the superstring. For the
elliptic or hyperbolic orbifolds, it was possible to extend the
orbifold construction to the superstring by choosing an appropriate
set of spin fields which were eigenfunctions of the momentum along the
compact direction, allowing us to construct superstring vertex
operators which satisfy the appropriate quantisation condition. We
were unable to find a corresponding basis for the parabolic orbifold
which gives the $M=0$ BTZ black hole; as a result, we cannot construct
superstring vertex operators which are well-defined on the orbifold
spacetime. This technical problem appears to be the most important
direction for further work: obviously, the main motivation for
interest in the $M=0$ and $M=J$ black holes is that they are
supersymmetric solutions in appropriate supergravity
theories~\cite{Coussaert:1993jp}. Also, the study of the tachyons in
twisted sectors is mainly interesting in the context of the
superstring, where we expect to be able to choose a GSO projection
which will project out the untwisted sector tachyon but keep the
tachyonic modes in odd twisted sectors. Any further progress on these
directions will require an appropriate construction of vertex
operators for the superstring.

\subsection*{Acknowledgements}
\label{acks}

We thank Mukund Rangamani for useful discussions. This work was
supported by the STFC. 

\bibliography{tachyon}
\bibliographystyle{utphys}

\end{document}